\documentclass[prl,twocolumn,amsmath,showkeys]{revtex4}

\usepackage{graphicx}
\usepackage{dcolumn}
\usepackage{bm}
\usepackage[T1]{fontenc}

\begin{document}

\title{Strong Isotopic Effect in Phase II of Dense Solid Hydrogen and Deuterium}

\author{Gr\'egory Geneste, Marc Torrent, Fran\c{c}ois Bottin, Paul Loubeyre}
\affiliation{CEA, DAM, DIF, F-91297 Arpajon, France}

\begin{abstract} 
Quantum nuclear zero-point motions in solid H$_2$ and D$_2$ under pressure are investigated at 80 K up to 160 GPa by first-principles path-integral molecular dynamics calculations. Molecular orientations are well-defined in phase II of D$_2$, while solid H$_2$ exhibits large and very asymmetric angular quantum fluctuations in this phase, with possible rotation in the (bc) plane, making it difficult to associate a well-identified single classical structure. The mechanism for the transition to phase III is also described. Existing structural data support this microscopic interpretation.
\end{abstract}

\keywords{hydrogen, deuterium, path-integral, quantum fluctuation, broken-symmetry phase}

\maketitle

Over the past 20 years, pressure-induced structural changes have been studied for most elements up to the 300 GPa range by a combination of synchrotron diamond anvil cell experiments and ab initio calculations. Although hydrogen has probably been the most extensively studied element under pressure, its crystalline
structures above 10 GPa are still largely unknown. Experimental structural determination is hampered by
hydrogen's weak x-ray scattering cross-section, and the use of neutron diffraction is equally, if not more,
difficult. Theoretically, the difficulties stem from the fact that nuclear zero-point motions and exchange contributions can have important effects. Novel quantum many-body states of matter should be expected, such as the recently predicted superconducting/superfluid metallic state \cite{babaev2004}. Understanding the phase diagram of hydrogen is thus an open challenge in modern condensed matter physics.

The phase diagrams of solid hydrogen and deuterium have been extensively
studied by Raman and infra-red vibrational spectroscopies. 
Phase boundaries have been determined for three phases\cite{mao94,cui95,goncharov11}. 
The transition from phase I to II appears to be related to a quantum ordering of the molecules, 
whereas that from II to III is more likely a classical ordering\cite{mazin97}. 
A strong difference in pressure has been measured between the I-II boundary line of solids H$_2$, HD and D$_2$, whereas there is almost none for the II-III phase boundary. 
X-ray diffraction studies have shown that the mass centers of the molecules remain on an hcp lattice at the I-II and II-III transitions, indicating that these phases are differentiated by a molecular ordering on this lattice\cite{goncharenko05,akahama10}. The nature of this ordering, however, is still unknown. 
Neutron diffraction data may rule out some theoretical predictions for phase II of
D$_2$ but are insufficient for determination of the orientational order\cite{goncharenko05}. 
Numerous calculations using Density Functional Theory (DFT) have tried to examine structural
changes in solid molecular hydrogen under pressure\cite{labet12}. 
A very small energy difference between various candidate structures is
obtained and the relative stability of these structures may be easily
changed by including zero-point energy contributions. The most recent
revision of the DFT phase diagram of hydrogen, including zero-point
energy at the harmonic level, predicts the $P6_3 /m$ structure for
phase II and a $C2/c$ structure for phase III \cite{pickard} though
neither of these structures is fully compatible with the experimental observations.

Correct treatment of the nuclear quantum effects involves
calculation of the zero-point motion of the protons (deuterons) beyond the harmonic
level and to include nuclear exchange to account for the differences
between ortho and para molecules. Taking into account the statistics
of the nuclei within an ab initio calculation is still a formidable
challenge. However, the nuclear quantum fluctuations can be accounted for exactly by a calculation that combines a DFT treatment of the electrons and a path-integral molecular dynamics (PIMD)
simulation of the nuclear motion. This technique (DFT-PIMD) was
applied to the case of hydrogen more than a decade ago in two
studies \cite{biermann1998,kitamura}. It was shown, in both
cases, that the quasi-harmonic treatment of the zero-point nuclear
motion is insufficient in high pressure solid hydrogen, and the two
studies produced different microscopic pictures of dense hydrogen. 
In one case, the quantum fluctuations of protons
effectively hinder molecular rotation in phase II \cite{kitamura} (as a quantum
localization) whereas in the other case, the structure was identified as
diffuse \cite{biermann1998} (coining the phrase "quantum fluxional
solid" (QFS)). 
Here we revisit a part of the DFT-PIMD phase diagram of hydrogen.
Nuclear zero-point motion is shown to have a strong influence in phase II:
the degree of orientational order is important in D$_2$, whereas
the structure of solid H$_2$ is found consistent with the concept of a QFS.

We have used the ABINIT code \cite{abinit} in which we have
implemented the path-integral formalism for nuclei. In this method,
the nuclei are treated quantum mechanically using a discrete
representation of the Feynman Path-Integral (PI) formulation of
quantum statistical mechanics. The PI formalism is based on an
isomorphism between the quantum system and a classical equivalent
system. Each quantum nucleus corresponds, in this classical
equivalent, to a ring polymer of $P$ classical nuclei, with
$P$ characterizing the discretization in imaginary time. 
The calculations were performed at 80 K. 
To determine the number of imaginary time slices $P$ (a compromise between accuracy and
computational burden), a test was performed on H$_2$ at 130 GPa (phase II), using $P$=64 and $P$=128. The angular density of probability obtained using 128 slices was found to be quasi-identical to that obtained using 64, thus $P$ was set to 64 in all the calculations.
In the strongly harmonic classical equivalent system, ergodicity is
efficiently recovered by using a Langevin thermostat
\cite{lan1,lan2,lan3} including a high-quality random number generator \cite{rand1}. 
For comparison, we have also computed classical trajectories by
setting $P$=1. The time step is 10 atomic time units ($\approx$ 0.24 fs) and 5 in the classical case. The mass centers of the molecules are initially distributed on an hexagonal lattice with the molecules pointing along the $c$ direction.

It should be noted that our calculation is not a full quantum treatment of the problem since we assume distinguishable nuclei (Botzmann statistics) and the decoupling
between nuclear and electronic degrees of freedom (Born-Oppenheimer
approximation). Moreover, we emphasize that the objective of the present study is not to
accurately determine phase transition pressures in H$_2$ and D$_2$ 
(which would require finite-size scaling analysis\cite{challa1986} difficult to achieve in ab initio), but rather to characterize molecular motions in phase II and its degree of orientational order. 
The DFT-PIMD computations were performed in the canonical ensemble, using the experimental lattice constants corresponding to pressures of 6, 30, 75, 100, 130,
145 and 160 GPa, in a 64-atom supercell for which the Brillouin Zone was sampled by 
a 2 $\times$ 2 $\times$ 2 $k$-point mesh (an orthorhombic supercell of size $2a \times 2
\sqrt{3} a \times 2c$ is constructed, $a$ and $c$ being taken from
experiments\cite{goncharenko05,akahama10,loubeyre96}). 
Imposing the lattice constants and a given periodicity is a strong constraint in a system where the different molecular structures are very close in energy\cite{pickard}.
However, even though the structure can be impacted, the relative effect -- from classical to D$_2$ to H$_2$ -- of quantum fluctuations on the molecular motions, and thus the degree of orientational order, should not be dependent on this constraint, and well reproduced by the canonical simulations.

For the sake of numerical efficiency, we have
introduced an additional level of parallelization in ABINIT
on the system replicas associated to the discretization of the PI in
imaginary time, beyond the already existing three levels of
parallelization (on k-points, bands and FFT grids
\cite{abinit-paral}). This level has a
quasi-linear scalability and, combined with the three others, allows
us to perform very long DFT-PIMD trajectories ($\geq$ 30 000 steps with a 64-atom supercell, and in excess of 100 000 steps in some cases where a high level of statistics is required). The first 5000 steps are used to equilibrate the system
and are thus excluded from the averages. The computations were
performed on the TERA-100 supercomputer at the CEA/DAM.
The density-functional treatment of electrons is achieved in the
Generalized Gradient Approximation (GGA) \cite{pbe}, using the
Projector Augmented-Wave (PAW) formalism \cite{paw}. GGA was
recently shown to give good agreement with experiments for the
equation of state \cite{caillabet10}. A plane-wave cut-off of 20
Hartree was required in the present study, in which the pressure
was limited to 160 GPa and hydrogen remains molecular and insulating.

\begin{figure}[htbp]
    {\par\centering
    {\scalebox{0.42}{\includegraphics{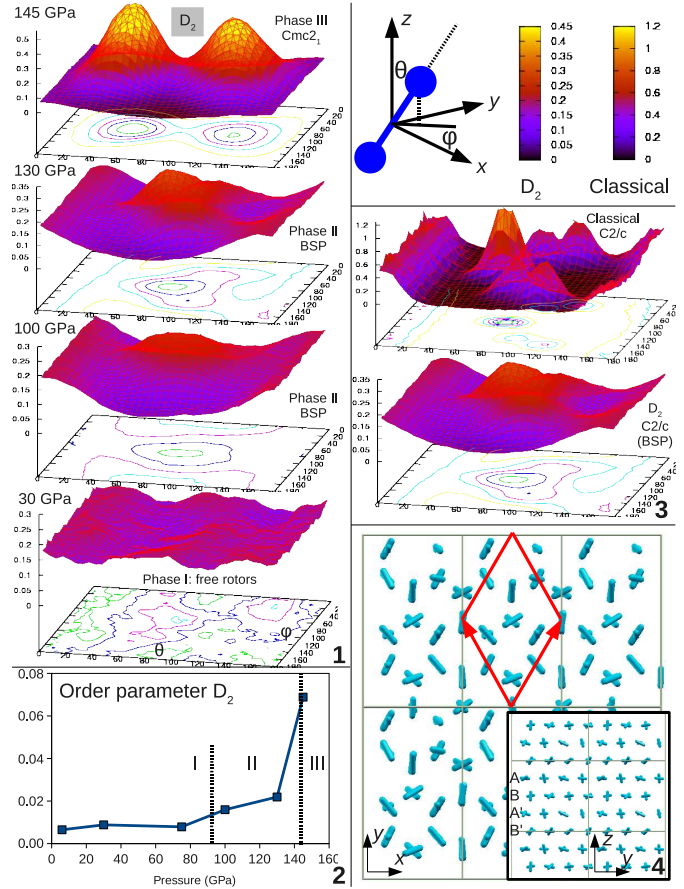}}}
    \par}
     \caption{{\small Structural properties of D$_2$ under pressure:
     (1) Pressure evolution of the ADoP: I (free rotors), II: BSP, III: $Cmc2_1$;
     (2) Pressure evolution of the order parameter;
     (3) ADoP of the classical and D$_2$ structures at 130 GPa, illustrating the effects of quantum fluctuations in D$_2$;
     (4) View of the structure of D$_2$ at 130 GPa (average over 20000 steps) -- red arrows: primitive cell.}}
    \label{figure1}
\end{figure}

The DFT-PIMD trajectories are analyzed using different outputs. We
define an order parameter according to \cite{cui1997}: $\tilde{O}
= < [\frac{1}{NP} \sum_{i=1}^{N} \sum_{s=1}^{P} P_2(\vec \Omega_{is}
. \vec e_y)]^2 >$, with $P_2(X) = \frac{1}{2}(3X^2 -1)$. The brackets $< ... >$
correspond to the average over the time steps, $N$ is the number of
molecules and $\vec \Omega_{is}$ is the vector of norm 1 lying along
the $i^{th}$ molecule of imaginary time slice $(s)$. The angular (normalized)
density of probability (ADoP) for the $(\theta,\phi)$ angles of the
molecular axis is obtained either by averaging over all of the
molecules of the supercell, giving  $f(\theta,\phi)$ or over
alternate families of planes along the $c$ axis, giving the (normalized) partial
angular densities of probability $f_i(\theta,\phi)$ ($i$=1,2).
$f_i(\theta,\phi) sin(\theta) \delta \theta \delta \phi$ is the
probality that a molecule of family $i$ points in the
$(\theta,\phi)$ direction with uncertainty $\delta \theta$ and
$\delta \phi$. The ADoPs are calculated by averaging over both PIMD
time steps and imaginary time slices, and therefore include both
thermal and quantum fluctuations. Finally, the structure is examined
by plotting the atomic positions, averaged over the real and
imaginary times of the simulation.

In the classical case, a disordered phase of free rotors is found
at low pressure while in the pressure range of phase II, a complex structure with a
32-atom primitive cell is found, identical to the $C2/c$ (idem $A2/a$) phase
described by Crespo {\it et al} \cite{crespo2011} (see Fig.~\ref{figure1}-4). 
This phase emerges from the Molecular Dynamics, and slightly differs 
from the $P6_3 /m$ structure found by the most recent DFT
calculations\cite{pickard} for phase II since it is of lower
symmetry (Fig.~\ref{figure1}-4). The $P6_3 /m$ structure has a 16-atom
unit cell with layers stacked in an ABAB fashion whereas the $C2/c$
structure has a 32-atom unit cell with layers stacked in the ABA'B'
fashion. Both AB and A'B' layers show a pinwheel motif. At high
pressure ($\approx$ 160 GPa), the $C2/c$ structure of the classical system
transforms into $Pca2_1$.

The structural evolution of solid D$_2$ is summarized in
Fig.~\ref{figure1}. The ADoP at different pressures are compared in
the first panel. At low pressure, the ADoP is almost
flat, similar to the one expected from the free rotor state.
With increasing pressure, clear contours appear in the angular distribution. 
Significant changes are seen between 75 and 100 GPa,
corresponding to the I-II transition. In addition, strong changes are
observed at 145 GPa, corresponding to the II-III transition. 
The evolution of the order parameter with pressure
(Fig.~\ref{figure1}-2) also shows these two phase transitions.
The ADoP of phase II in D$_2$ is compared to that of classical
hydrogen at 130 GPa in Fig.~\ref{figure1}-3. 
The classical and D$_2$ ADoP look similar, but in the
case of D$_2$ the ADoP is considerably smoothed by the quantum fluctuations. 
However, despite these strong fluctuations, the atomic positions in the supercell
yield a configuration close to the $C2/c$ structure (Fig.~\ref{figure1}-4).
The important point is that in this broken-symmetry phase (BSP) of D$_2$ at 130 GPa,
the molecular orientations are well-defined, making it possible 
to assign a classical structure in this pressure range.
At 145 GPa, phase II discontinuously transforms into a $Cmc2_1$
structure (phase III) with molecules of each basal plane having the same orientation, 
alternating along the $c$ axis. All the calculations with pressures < 160 GPa exhibit phases in which the molecules remain on an hcp lattice, in agreement with the experiments.

\begin{figure}[htbp]
    {\par\centering
    {\scalebox{0.4}{\includegraphics{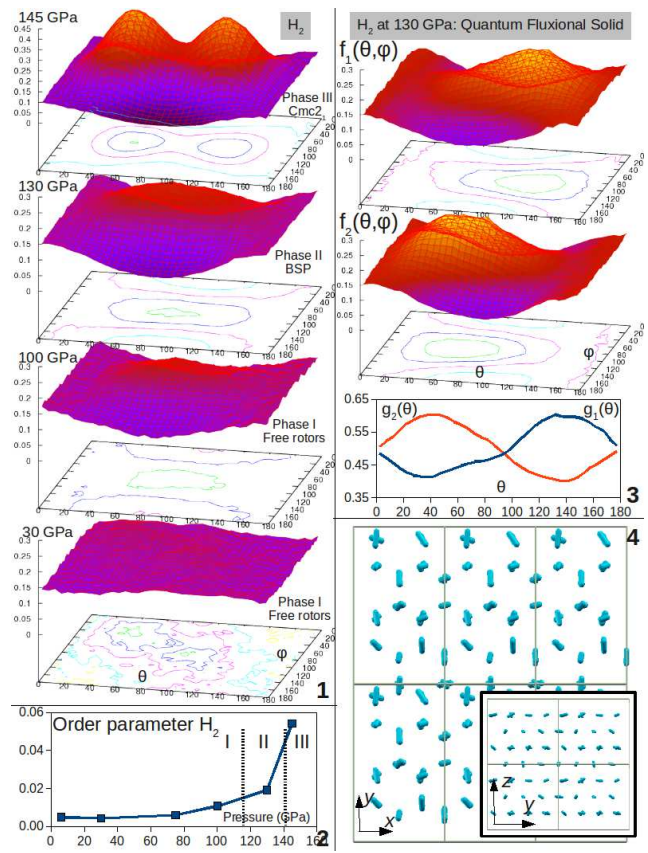}}}
    \par}
     \caption{{\small Structural properties of H$_2$ under pressure:
     (1) Pressure evolution of the ADoP: I (free rotors), II: BSP, III: $Cmc2_1$;
     (2) Pressure evolution of the the order parameter;
     (3) QFS at 130 GPa: f$_1$, f$_2$ and integration over $\phi$ of these functions, as a function of $\theta$;
     (4) View of the structure at 130 GPa (average over 20000 steps).}}
    \label{figure2}
\end{figure}

In the case of hydrogen, the stronger quantum effects yield a surprisingly
different microscopic picture (Fig.~\ref{figure2}). At low pressure,
the free rotor phase is associated with a flat ADoP. Upon increasing
pressure, the molecules start to preferentially move in the (bc)
plane ($\phi$ = 90$^{\circ}$), as evidenced by the progressive
appearance of a broad maximum around this value
(Fig.~\ref{figure2}-1). Relying only on the ADoP $f(\theta, \phi)$, it is
difficult to detect a symmetry breaking until 145 GPa. At this pressure,
clear changes are observed: the ADoP exhibits two peaks at $\theta_{max}$ $\approx$ 52.5
and 127.5$^{\circ}$ ($< \theta >$ = 63 and 117$^{\circ}$), 
with angular width $(\Delta \theta, \Delta \phi) \approx$  (28$^{\circ}$,
37$^{\circ}$): hydrogen evolves towards a well-defined $Cmc2_1$ structure similar to the one
found in Ref.~\onlinecite{kitamura} at similar pressures. 
As expected, the quantum fluctuations in this $Cmc2_1$ phase are 
larger than in $Cmc2_1$ D$_2$ at the same pressure.
On the other hand, no clear discontinuity is observed in the
evolution of the order parameter below this pressure
(Fig.~\ref{figure2}-2).

In fact the symmetry breaking associated with phase II occurs at a
lower pressure: it is not directly apparent in the ADoP $f(\theta,
\phi)$, but in $f_1$ and $f_2$ and their $\phi$-integrated counterparts $g_1(\theta)$ and $g_2(\theta)$
(Fig.~\ref{figure2}-3). Below 130 GPa these are quasi-identical
but at 130 GPa, $f_1$ and $f_2$ differ significantly. $f_1$ has a
broad maximum at $(\theta,\phi) \approx$ (122.5$^{\circ}$,
90$^{\circ}$) while $f_2$ has a broad maximum at $(\theta,\phi)
\approx$  (57.5$^{\circ}$, 90$^{\circ}$), reflecting a symmetry
breaking along the $c$ axis that is not evident in the global ADoP $f$, which remains single-peaked. 
The solid simulated at 130 GPa is thus in phase II, while the well-ordered structure at 145 GPa ($Cmc2_1$)
is in phase III. The atomic positions in phase II, obtained by an
average over 20000 steps, are shown in Fig.~\ref{figure2}-4. No
clear structure emerges, reflecting the very large quantum
fluctuations of protons in phase II. For $f_1$ and $f_2$, $(\Delta
\theta, \Delta \phi) \approx$ (39$^{\circ}$, 45$^{\circ})$. The
fluctuations in $\theta$ are also very asymmetric, as shown by the
quantum mechanically averaged values of $(\theta,\phi)$ $\approx$
(85$^{\circ}$, 90$^{\circ}$) and (95$^{\circ}$, 90$^{\circ}$), which
are very different from those of the orientation of maximal probability
($\theta_{max}$ = 122.5 and 57.5$^{\circ}$). Additional calculations 
using a 256-atom supercell or $P$=128 at such pressure provide similar
results.

The averaged structure at 130 GPa is thus close to $Cmcm$ -- a phase with the molecules lying in the (ab) plane -- whereas the structure corresponding to the maximum of angular probability would be rather the $Cmc2_1$ type. This difference is due to large and strongly asymmetric angular quantum fluctuations. 
Two different structures may thus be predicted depending on whether the most probable molecular orientations or the quantum mechanically averaged ones are considered.
Moreover, inspecting the ADoP of each molecule reveals that
they can, to some extent, rotate in the (bc) plane, showing that phase II of H$_2$ cannot be modeled at the harmonic level. A few years ago, Biermann {\it et al} \cite{biermann1998} introduced the notion of "quantum fluxional solid" to qualify a solid for which an underlying
classical structure cannot be clearly identified because of large
quantum fluctuations. We suggest this is precisely the case for phase II of
hydrogen.

Upon entering phase III, H$_2$ and D$_2$ adopt the $Cmc2_1$
structure. At pressures $\ge$ 160 GPa, the structure in both solid isotopes evolve towards a $Cmca$ space
group, in which the molecular centers are no longer on the hcp lattice, showing the limit of stability for structures in which the molecules remain on the hcp sites. However a small increase in pressure should result in a displacive phase transition. Examination of the nature of phase III is precluded at higher pressures as the molecular centers move away from an hcp configuration and is thus beyond the scope of the present work.

\begin{figure}[htbp]
    {\par\centering
    {\scalebox{0.28}{\includegraphics{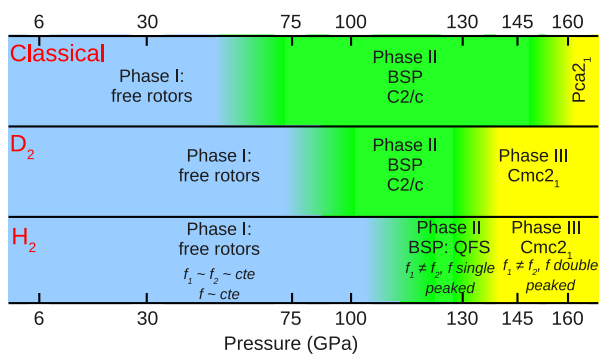}}}
    \par}
     \caption{{\small Proposed pressure evolution of dense hydrogens at T=80 K.}}
    \label{figure3}
\end{figure}

The proposed phase diagrams at 80 K are summarized in Fig.~\ref{figure3} for H$_2$, D$_2$ and the classical case. The most important point is that the molecular orientations are relatively well-defined in the phase II of D$_2$, which is not the case in the phase II of H$_2$, where the concept of QFS can be used.
There is a strong shift in the I-II transition pressure from D$_2$ to H$_2$, in quantitative agreement with experiment. 
In contrast to the DFT-PIMD calculation of Ref.~\onlinecite{kitamura}, no quantum localization is
observed: quantum effects systematically induce a gain in symmetry (favor molecular rotation). 
Experimentally, a superstructure peak is observed in phase II along the $a$ axis for
D$_2$\cite{goncharenko05} but along the $c$ axis for H$_2$\cite{loubeyre96}. 
The supercell volume was not large enough to permit observation of such superstructures in the
present simulations. Nevertheless, such superstructure peaks indicate that the symmetry breaking is dominant
along the $c$ axis for solid H$_2$ and in the (ab) plane for solid D$_2$, as suggested here. 
The very small intensity difference in the neutron diffraction peaks measured for solid D$_2$ at the I-II transition can be explained by the very small difference in the broad ADoP and order parameter at this transition. Since the change is even weaker in the case of solid H$_2$, the structural refinement of
phase II appears to be beyond the reach of current experiments. 
The II-III transition pressure is very similar for both isotopes, between 130 and 145 GPa, in qualitative agreement with experiment (152-165 GPa\cite{goncharov11}).
Future studies will extend the present calculation over the pressure domain of phase III 
in the isothermal-isobaric ensemble, thus excluding any assumption regarding the lattice constants (unlike in the present work).

\end{document}